\documentclass[aps,pre,twocolumn,superscriptaddress]{revtex4}
\usepackage{graphicx}
\usepackage{amsmath}

\begin{document}

\title{Design of an Externally Driven Current Cloak}



\author{A. S. Garc\'{i}a-Gordillo}
\affiliation{Group of Complex Systems and Statistical Physics, Physics Faculty, University of Havana, 10400 Havana, Cuba}

\author{E. Altshuler}
\email[]{ealtshuler@fisica.uh.cu}
\affiliation{Group of Complex Systems and Statistical Physics, Physics Faculty, University of Havana, 10400 Havana, Cuba}


\date{\today}

\begin{abstract}
An inhomogeneity into a conductive matrix deforms the flow pattern of an applied electric current. A usual current cloak can be defined as a permanent modification of the matrix properties around the inhomogeneity guaranteeing that the current flow pattern is similar before and after passing by the modified zone, so it implies the ``electrical invisibility" of the inhomogeneous region. Here we introduce the concept of a current cloak that can be tuned --switched on and off, for example-- by means on an external field. We demonstrate analytically and using Finite Elements Simulations that a current cloak can be constructed and manipulated by an external magnetic field for a concrete system consisting in a magneto-resistive matrix with a stainless steel inclusion.
\end{abstract}

\pacs{72.80.Tm, 81.05.Ni, 72.15.Gd}
\maketitle

\section{Introduction}

Lack of homogeneity in materials is generally assumed as deleterious: uncontrolled pores, second phase inclusions, grain boundaries and many other defects tend to weaken mechanical, electrical and magnetic properties \cite{Frankel1957,Altshuler1999,Batista1999,Batista2003_resistive,Batista2003,Mo2013}. In fact, one of the main aims of Materials Science is to improve fabrication methods in order to eliminate them.

One alternative approach to the problem is not eliminating, but concealing these defects: this is at the core of the emerging field of {\it cloaks}. They can be defined as materials modified in such a way that an inhomogeneous region inside them is made invisible (i.e., indistinguishable from its surroundings) to electromagnetic fields or sound waves, for example. The subject was kicked off by the pioneering theoretical work of Pendry and Leohardt in {\it optical} cloaks \cite{Pendry2006,Leonhardt2006}: if a region of space has optical properties different from the rest of the material, it is possible to engineer a change of properties in the vicinity of the inhomogeneity so the rays of light detected from a point far from the region follow the same trajectories they had before entering the region in question. Pendry and Leohardt coined the term {\it metamaterial} to describe a material engineered in such way. Over the following years, several types of cloaks were reported in the optical \cite{Valentine2009,Gabrielli2009,Smolyaninov2009,Ergin2010,Rama2015} and microwave \cite{Schurig2006,Liu2009,Tretyakov2009,Edwards2009,Ma2010} frequency ranges.

Later on, cloaks invaded the scenario of stationary (or near-stationary) fields. Wood and Pendry were the first to propose a cloak to conceal a defect in the presence of a low-frequency magnetic field \cite{Wood2007}. The idea was experimentally proven a few years later by a combination of two materials with ``opposite" magnetic behaviours: a superconductor and a ferromagnet \cite{Narayana2012}. The boundary was immediately pushed further to include stationary fields \cite{Gomory2012}. Yungui Ma {\it et al.} designed and obtained experimentally a ``bifunctional cloak" able to conceal the lines of electrical current ({\it current cloak}) or heat flux lines ({\it heat cloak}) around a cavity in a metal, or an inclusion with conductivity different from its surroundings \cite{Ma2014}. Importantly, in all the cloaks described above the material properties must be permanently modified  around the inhomogeneity in order to achieve the desired effect.

Here, we propose a new way to achieve the cloaking effect: instead of permanently modifying the material, an external field is used to ``tune" or ``drive" the material properties around the inhomogeneity. We illustrate the idea with a current cloak able to suppress the deformation of the current lines around an inclusion of larger conductivity within a magneto-resistive matrix. Firstly, the feasibility of achieving the cloaking effect is demonstrated analytically. Then, we show it by means of Finite Elements Simulations for a realistic set of materials and magnetic fields. We call the new metamaterial Externally Driven Current Cloak or, for our particular case, Magnetically Driven Current Cloak. Finally, we briefly examine the interaction of cloaks between them, and with a boundary.

\section{The magnetically driven current cloak}

\subsection{General description}

\begin{figure}[htb]
\begin{center}
\includegraphics[width=0.70\linewidth]{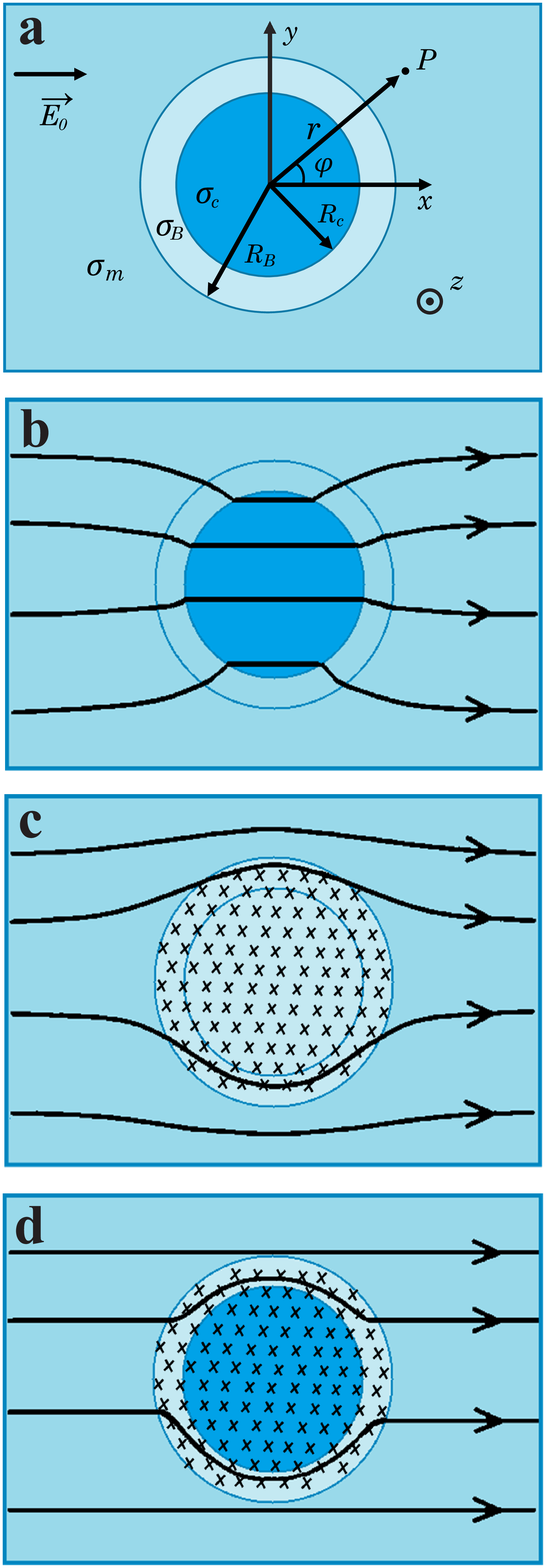}
\caption{\label{FIG:Sketches} A two-dimensional current cloak driven by an external magnetic field. (a) Parameters used to derive the cloaking condition (see text). (b) Magneto-resistive matrix with inclusion at zero magnetic field.  (c) Matrix alone at cloaking magnetic field, $B_{cloak}$ (d) Matrix with inclusion at $B_{cloak}$. In (b-d), darker blue indicates larger conductivity, black lines represent the electric current (or electric field) lines. In (c and d), the applied magnetic field is represented by ``x''s.}
\end{center}
\end{figure}

Fig. \ref{FIG:Sketches}(a) shows a sketch of a 2D version of the magnetic field driven current cloak. The inner circle of radius $R_c$ is the inclusion (or core), whose conductivity $\sigma_c$ is larger than that of the magneto-resistive matrix at zero applied field, $\sigma_m$. The circle of radius $R_B$ is the area covered by a magnetic field $B$ which is applied in a direction perpendicular to the plane of the matrix. As a result, the conductivity increases from $\sigma_m$ to $\sigma_B$ within a ring of inner and outer radii $R_c$ and $R_B$, respectively.

Fig. \ref{FIG:Sketches} (b)-(d) illustrates the working principle of the cloak: the inner and outer circumferences have radii $R_c$ and $R_B$ respectively, which are not explicitly indicated for clarity; the black lines correspond to an electric field (or current) which is uniform far from the core. Fig. \ref{FIG:Sketches} (b) shows the matrix and the core at zero magnetic field: the field lines ``bend towards" the core, due to the fact that $\sigma_{c} > \sigma_{m}$. Fig. \ref{FIG:Sketches} (c) corresponds to the system at a magnetic field $B_{cloak}$ applied within the region of radius $R_B$, without showing the effect of the core: due to the decrease of the conductivity in the magneto-resistive ring, the current lines ``bend away" from the core --an effect that compensates that described in (b). Fig. \ref{FIG:Sketches} (d) illustrates the effect of $B_{cloak}$ on the matrix-core composite, so the flow lines are the overlapping between the patterns sketched in (b) and (c): the result is that the current flow becomes homogeneous immediately outside the region containing the magnetic field --the current cloak effect has been achieved. Now, we calculate the magnetic field needed to produce the cloaking,  $B_{cloak}$, as a function of the radii and conductivities.

\subsection{An analytical expression for the cloaking condition}

The continuity equation for an electric current density $\vec{J}$, represented by the black lines in Fig. \ref{FIG:Sketches} reads $\vec{\nabla} \cdot \vec{J} = 0$. If $\vec{E}$, $V$ and $\sigma$ are the electric field, the potential and the conductivity, one can write $\vec{J} = \sigma \vec{E}$ (where $\vec{E}=-\vec{\nabla}V$). By combining the previous expressions, we get a Laplace equation for the potential $V$ in cylindrical coordinates.

If we solve it for the system illustrated in Fig. \ref{FIG:Sketches}(a) and impose the cloaking condition, i.e., $\vec{E}$ is uniform outside the area where the magnetic field is applied, we get (see Supporting Online Material):
  \begin{equation} \label{cloaking_condition}
 \sigma_m= \frac{R_B^2(\sigma_c+\sigma_B)+R_c^2(\sigma_c-\sigma_B)}{R_B^2(\sigma_c+\sigma_B)+R_c^2(\sigma_B-\sigma_c)} \sigma_B
  \end{equation}

It is worth noting that the cloaking condition reported by Ma {\it et al.} for a core consisting in a void \cite{Ma2014} is a particular case of Eq. \ref{cloaking_condition} corresponding to $\sigma_c = 0$. Both conditions stand for any value of applied electric field, i.e., no matter the value the uniform current injected through the sample takes, the cloaking parameters will remain the same.

If we follow the thread of previous work in the field, the cloaking condition implies inserting a ring of conductivity $\sigma_B$ and inner and outer radii $R_c$ and $R_B$, respectively, so all the parameters fulfill Eq. \ref{cloaking_condition}. Our approach is different:  since $\sigma_B$ is a function of the magnetic field, we propose applying a magnetic field B$_{cloak}$ on the ring-shaped region in such a way that the resulting conductivity matches the value given by Eq. \ref{cloaking_condition}. Naturally, $B_{cloak}$ depends on the magneto-resistive behaviour of the matrix.

\subsection{A realistic magnetically driven current cloak}

\begin{figure}[b]  
\begin{center}
\includegraphics[width=0.95\linewidth]{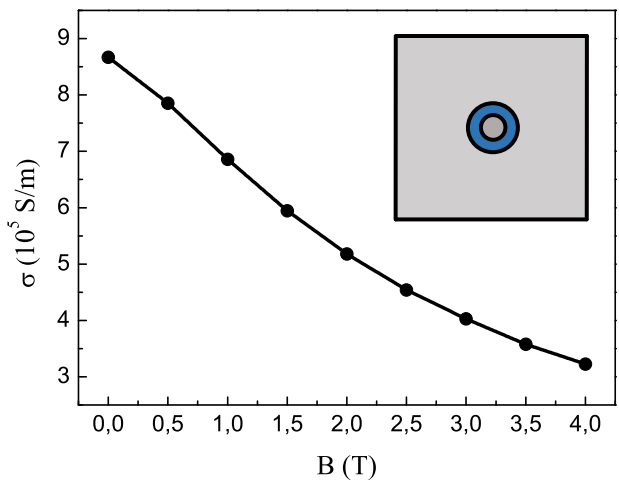}
\caption{\label{FIG:Bismuth} Magnetic field dependence of conductivity for polycrystalline Bi (data taken from \cite{Yang1999}). }
\end{center}
\end{figure}

Now, we analyze the case of a composite made of a polycrystalline Bi matrix shaped as a 25 $\times$ 25 $\times$ 1 mm$^3$ sheet, whose magneto-resistive behavior is illustrated in Fig. \ref{FIG:Bismuth} (8.67 $\times$ 10$^5$ S/m at zero field \cite{Yang1999}). A cylindrical inclusion of stainless steel of radius 1.6 mm is located at the center of the matrix. The conductivity of the inclusion --which is very weakly dependent on the magnetic field-- is taken as 9.93 $\times$ 10$^6$ S/m \cite{Conductivities}. The external field is confined to a region of radius 3.2 mm around the inclusion, which can be easily achieved in practice by using a couple of cylindrical super-magnets of appropriate radii.

\begin{figure}[t]  
\begin{center}
\includegraphics[height=16.48227cm, width=7.5cm]{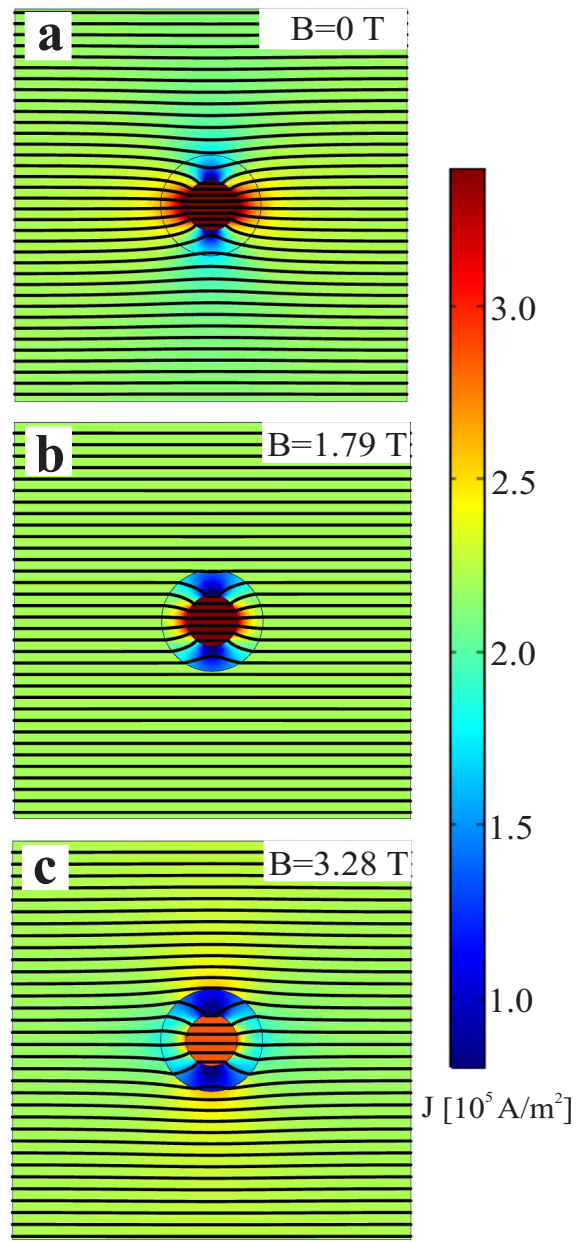} 
\caption{\label{FIG:FEMsimul} Magnetically driven cloaking for a realistic sample. FEM simulations of the current density lines for a 1-mm thickness polycrystalline Bi matrix with a stainless steel inclusion of $R_c =$ 1.6 mm radius (inner circle), and a magnetic field applied within a region of radius $R_B =$ 3.2 mm (outer circle). (a), (b) and (c) correspond to magnetic fields of $0$ T, $1.79$ T and $3.28$ T, respectively. Here, $B_{cloak}$ $\approx$ 1.79 T.}
\end{center}
\end{figure}

Fig. \ref{FIG:FEMsimul} shows FEM simulations of the electric current density lines along the sample at different magnetic fields perpendicular to the large face of the matrix (a current of $5.5$ A is uniformly injected along the left vertical boundary of the matrix with a uniform density $J_{in}$ $=$ $2.2$ $\times$ 10$^5$ A/m$^2$ in the case of no inclusion and zero applied magnetic field). For zero applied field (Fig. \ref{FIG:FEMsimul} (a)) the lines are deformed towards the inclusion, corresponding to the case sketched in Fig. \ref{FIG:Sketches} (b). For a field $B$ = $1.79$ T (Fig. \ref{FIG:FEMsimul} (b)), the deformation of the current lines towards the highly conductive core is compensated by the ring of depleted conductivity due to the effect of the applied magnetic field, so they are homogeneous outside the zone where the magnetic field is confined. This situation, analogous to the one sketched in Fig. \ref{FIG:Sketches} (d) implies that $B_{cloak}$ $\approx$ $1.79$ T for our composite. In fact, this is the field value that satisfies Eq. \ref{cloaking_condition}, taking for the specific geometry and conductivities of the cloak, as well as the curve shown in Fig. \ref{FIG:Bismuth}. For a magnetic field of $3.28$ T (Fig. \ref{FIG:FEMsimul}(c)), the conductivity of the ring around the inclusion has decreased to a point that it ``over-compensates" the original deformation of the current lines.

\begin{figure}[b]
\begin{center}
\includegraphics[width=0.95\linewidth]{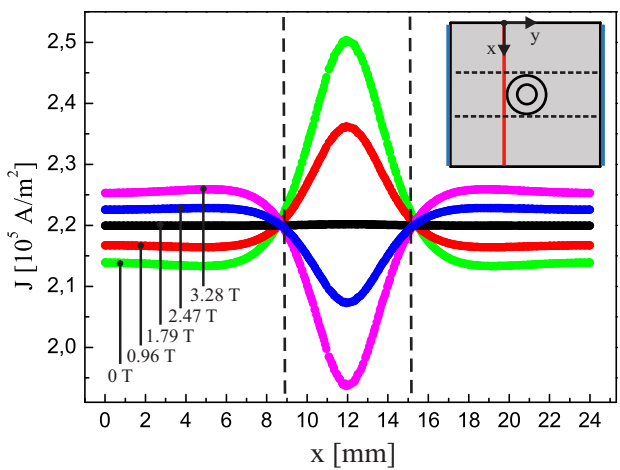}
\caption{\label{FIG:CloakCriterium} Current density values plotted along a vertical line tangent to the area of application of the magnetic field (see inset). The straightest curve ($B_{cloak}$ = 1.79 T) fulfills the cloaking condition.}
\end{center}
\end{figure}

In order to quantify the current density distribution near the cloak, and properly estimate $B_{cloak}$, a more detailed analysis is needed. Fig. \ref{FIG:CloakCriterium} shows how the electric current density changes along a vertical line running from the top to the bottom of the sample, which is tangent to the region where the magnetic field is applied (see inset of Fig. \ref{FIG:CloakCriterium}). The two vertical dotted lines in the main graph correspond to the upper and lower edges of the region where the magnetic field is applied, respectively. For $B$ = $0$, the current density above and below that region decreases from its value when there is no inclusion nor magnetic field applied, since the current lines ``bend towards" the higher conductivity core. The bending is smaller for $B$ = $0.96$ T and virtually non-existent for $B$ = $1.79$ T. The lines bend in the opposite direction for fields of $2.47$ T and $3.28$ T. In the region where the magnetic field is applied, along the same vertical line, the deformation of the originally uniform current density field is more evident, even so, for $B$ = $1.79$ T there is no tangible sign of deformation. This magnetic field was chosen according to Eq. \ref{cloaking_condition} for the cloaking condition to be fulfilled. It should be pointed out that $B_{cloak}$ does not change if the sample is larger in the {\it x} direction (see reference frame of Fig. \ref{FIG:Sketches} (a)); as a matter of fact, the current density profiles of Fig. \ref{FIG:CloakCriterium} also remain the same as expected.

Based on the curves shown in Fig. \ref{FIG:CloakCriterium}, the deviations of the current density from its uniform value when there is no inclusion nor magnetic field applied can be quantitatively evaluated by defining a cloaking parameter:

\begin{equation} \label{cloaking_parameter}
 P_{cloak}= 1 - \frac{Max(|J(x)-J_{in}|)}{J_{in}}
  \end{equation}

\noindent where $J_{in}$ is the uniform current density for no inclusion or magnetic field applied and $J(x)$ is the current density dependence along the vertical line displayed in the inset of Fig. \ref{FIG:CloakCriterium} for a certain external magnetic field and $R_B$. This allows us to quantify how far we are from achieving the cloak. If the applied magnetic field is different from $B_{cloak}$, then $P_{cloak}<1$, suggesting that the device is not in the cloaking state. On the other hand, if the applied field coincides with $B_{cloak}$, then $P_{cloak}=1$ and the device will behave as a genuine current cloak.

\begin{figure}[t]
\begin{center}
\includegraphics[width=0.97\linewidth]{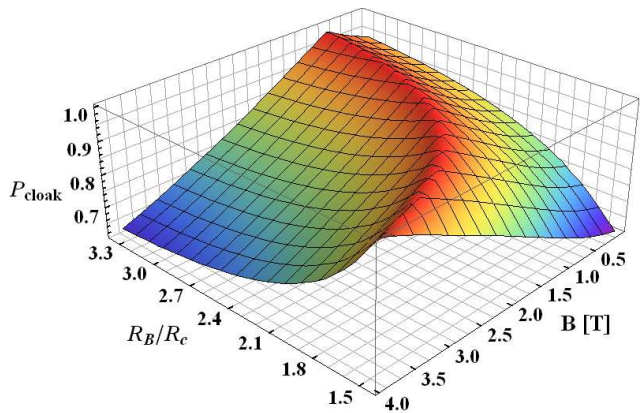}
\caption{\label{FIG:3D} Exploring the cloaking parameter space. Dependance of $P_{cloak}$ with $R_B$ and the external applied magnetic field $B$: cloaking and non-cloaking states.}
\end{center}
\end{figure}

By varying the area of the zone where the external magnetic field is applied (by modifying its radius $R_B$) and then varying the value of the applied magnetic field in each previous configuration, it is possible to build a ``parameter phase diagram'' using the definition of $P_{cloak}$ given by Eq. \ref{cloaking_parameter}. Fig. \ref{FIG:3D} illustrates the behavior of $P_{cloak}$ with $R_B/R_c$ and the applied magnetic field $B$. These magnitudes were chosen because they can be externally modified to achieve the cloaking condition for a given inclusion into a given matrix. The cloaking and non-cloaking states are visible in Fig. \ref{FIG:3D} in terms of $P_{cloak}$: the red ridge corresponds to the sets of external parameters that make possible the cloaking. The smaller the radius $R_B$ the strongest the magnetic field we have to apply to achieve the current cloaking condition. These sets of parameters, for which $P_{cloak}=1$, match those calculated using Eq. \ref{cloaking_condition} within negligible uncertainty ($\approx$0.1$\%$). Although the red ridge follows Eq. \ref{cloaking_condition} for a given magneto-resistive matrix, an analytical expression for the specific cloaking state given by the rest of the parameters in Fig. \ref{FIG:3D} is not easy to obtain.

Fig. 5 also illustrates that the cloaking condition is achievable even for $R_B$ values comparable to the size of the matrix, suggesting that its boundaries do not affect considerably the cloaking effect. This is shown in the Supporting Online Material.

Up to now we have only considered the modification of the current flow pattern by the presence of one inhomogeneity in our magneto-resistive matrix. If two identical inclusions deployed side-by-side along the symmetry axis of the sample along the y-axis, the cloaking affect is again achieved for the same value of the magnetic field (see Supporting Online Material). This is presumably the case for a larger number of inclusions, provided the matrix is long enough along the y-axis.

In summary, we have proposed the idea that current cloaks can be tuned externally. We demonstrate that the cloaking condition can be achieved in the case of a circular inclusion embedded in a magneto-resistive plate, by applying a perpendicular magnetic field of appropriate intensity. We provide enough quantitative information to attempt an experimental test of the concept. It is reasonable to believe that other kinds of externally driven currents cloaks (based, for example, on temperature gradients) can be conceived and materialized.

\section {Acknowledgements}
We acknoweledge Adri\'{a}n Enrique for suggesting the possibility of using a magneto-resistive material for the matrix and Y. Nahmad for letting us use COMSOL Multiphysics for the simulations. This research was partially supported by the University of Havana's Institutional Project ``Superconductors and conductors: from characterization to applications". E. A. drew
inspiration from the late M. \'{A}lvarez-Ponte.

\appendix
\section{Supporting Online Material}\label{Supporting Online Material}

\subsection{An analytical expression for the cloaking condition}

\begin{figure}[t]
\begin{center}
\includegraphics[width=0.7\linewidth]{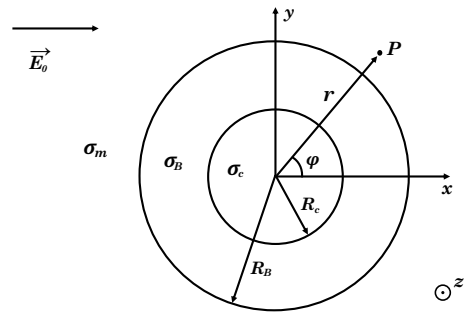}
\caption{\label{FIG:scheme analitic solution} Sketch of the magnetically driven current cloak. Parameters used to derive the cloaking condition (see text).}
\end{center}
\end{figure}

We consider a cylindrical inclusion (infinite along {\it z}) of radius $R_c$, whose conductivity $\sigma_c$ is
larger than that of the magneto-resistive matrix $\sigma_m$ where it is immersed. In a concentric region of radius $R_B$ an external magnetic field is applied. Due to the presence of this field the conductivity of a hollow cylinder of inner and outer radii $R_c$ and $R_B$ changes to $\sigma_B$, being $\sigma_B$ smaller than $\sigma_m$ for any non-zero magnetic field values. The electric field at infinity is shown in Fig. \ref{FIG:scheme analitic solution} with a sketch of the system.

To deal with an electric stationary problem like the current cloak we used the current density continuity equation:

\begin{equation}\label{continuity1}
 \textbf{$\vec{\nabla}$} \cdot \textbf{$\vec{J}$}=0
 \end{equation}

 \noindent considering that $\vec{J}=\sigma \vec{E}$, $\vec{E}=-\vec{\nabla} V$, and the divergence of the gradient of a function is equal to the laplacian of itself, we get:

\begin{equation}\label{continuity2}
 \begin{split}
 \textbf{$\vec{\nabla}$} \cdot (\sigma \textbf{$\vec{E}$})=0 \\
 -\textbf{$\vec{\nabla}$} \cdot \sigma \textbf{$\vec{\nabla}$}V=0 \\
 \sigma \textbf{$\nabla^2$}V=0  \\  \textbf{$\nabla^2$}V=0
 \end{split}
 \end{equation}

For the present system it is suitable to use cylindrical coordinates. The Laplace equation for the potential in this coordinate system reads:

 \begin{equation} \label{laplaciano_cilindricas}
 \frac{1}{r} \frac{\partial}{\partial r} (r \frac{\partial V}{\partial r}) + \frac{1}{r^2} \frac{\partial^2 V}{\partial \varphi^2} + \frac{\partial^2 V}{\partial z^2} =0
 \end{equation}

If we take into account the geometry of the problem in the way that the lines of $\vec{E}$ (parallels to the lines of $\vec{J}$) come from $-\infty$ in the {\it x} axis -orthogonal direction to the {\it z} axis- (see Fig. \ref{FIG:scheme analitic solution}) it is not difficult to realize that:

 \begin{equation} \label{no_depende_de_z}
 V(r ,\varphi, z)= V(r ,\varphi)
 \end{equation}

 \noindent hence:

 \begin{equation} \label{z_derivada}
 \frac{\partial^2 V}{\partial z^2} =0
 \end{equation}

This independence regarding the {\it z} coordinate allows us to treat as equivalent three qualitative different systems: the one which is infinite in the {\it z} direction, the one with finite {\it z} dimensions and the $2D$ system. The case presented in this paper, due to practical implications, was the system with finite dimensions consisting in a foil with an inclusion resembling a coin and a magnetic field affected washer around it.

If we propose the solution $V(r,\varphi)=R(r) \cos \varphi$ and substitutes it in Eq \ref{laplaciano_cilindricas} (without the {\it z} related term), the expression reduces to an Euler's equation for $R(r)$:

\begin{equation} \label{euler equation}
 r^2 \frac{\partial^2 R}{\partial r^2} + r \frac{\partial R}{\partial r} - R = 0
 \end{equation}

Proposing the classic solution to this equation as $R=r^k$, we get the coefficients $k=\pm 1$. Combining the two solution proposals we get the following expressions for the potential in the three regions of interest:

\begin{equation} \label{solution}
   V(r ,\varphi)=\begin{cases}
                 \big(C_1 r + \frac{C_2}{r}\big) \cos \varphi \textbf{ } \textbf{ } \textbf{ }\textbf{ }\textbf{ } \textbf{ } r>R_B \\
                 \big(C_3 r +\frac{C_4}{r}\big) \cos \varphi \textbf{ } \textbf{ } \textbf{ }\textbf{ }\textbf{ } \textbf{ } R_c<r<R_B \\
                 \big(C_5 r +\frac{C_6}{r}\big) \cos \varphi \textbf{ } \textbf{ } \textbf{ }\textbf{ }\textbf{ } \textbf{ } r<R_c
                 \end{cases}
\end{equation}

\noindent where $C_1$, $C_2$, $C_3$, $C_4$, $C_5$ and $C_6$ are constants to be calculated from the boundary conditions.

The expression for the potential at the infinity considering an uniform electric field of intensity $E_0$ parallel to the {\it x} axis reads:

 \begin{equation} \label{condicion_en_infinito}
 V(\infty,\varphi,z)= -E_{0} r \cos \varphi
 \end{equation}

Moreover, the potential at $r=0$ cannot reach infinite, so it must be bounded:

 \begin{equation} \label{condicion_en_cero}
 \mid V(0 ,\varphi)\mid < \infty
  \end{equation}

From the expressions \ref{condicion_en_infinito} and \ref{condicion_en_cero}, and the solution \ref{solution} for $V(r ,\varphi)$ we get:

 \begin{equation} \label{C1}
 C_1=-E_0
  \end{equation}

and

 \begin{equation} \label{C6}
 C_6=0
  \end{equation}

\noindent remaining only four constants to solve.

In order to calculate these constants we need to look at the boundary conditions at $R_c$ and $R_B$. In each of these boundaries stands the continuity of the potential and the continuity of the normal component of the current density ($\vec{J}$):

\begin{equation}\label{condiciones_de_frontera_en_R1_y_R2}
 \begin{split}
 V_B(R_c,\varphi)=V_c(R_c,\varphi) \\
 V_m(R_B,\varphi)=V_B(R_B,\varphi) \\
 J_{Bn}(R_c,\varphi)=J_{cn}(R_c,\varphi) \\
 J_{mn}(R_B,\varphi)=J_{Bn}(R_B,\varphi)
 \end{split}
 \end{equation}

Substituting the potentials and the normal components of the current density ($J_n$) evaluated at $R_c$ y $R_B$, and taking into account the gradient of $V$ in cylindrical coordinates for which:

\begin{figure}[b]
\begin{center}
\includegraphics[width=0.7\linewidth]{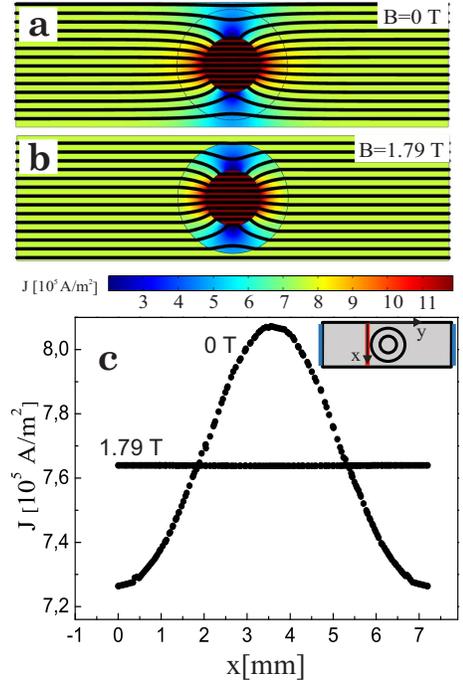}
\caption{\label{FIG:boundariesproximity} Proximity of the boundaries. (a,b) Current density lines for a 1-mm thickness polycrystalline Bi matrix with a stainless steel inclusion of $R_c =$ 1.6 mm radius (inner circle), and a magnetic field applied within a region of radius $R_B =$ 3.2 mm (outer circle). Upper and lower boundary positions were changed allowing a separation of $1/7$ its radius from the device. (a) and (b) correspond to magnetic fields of $0$ T and $1.79$ T, respectively. (c) Current density values along a vertical line tangent to the area of application of the magnetic field (see inset).}
\end{center}
\end{figure}

\begin{equation} \label{Electric field en cilindricas}
 \vec{E}= -\frac{\partial V}{\partial r} \hat{r} - \frac{1}{r} \frac{\partial V}{\partial \varphi} \hat{\varphi}
  \end{equation}

\noindent we obtained the following system of equations in four variables:

\begin{equation}\label{sistema de 4x4}
 \begin{split}
 C_3 R_c + \frac{C_4}{R_c}= C_5 R_c \\
 -E_0 R_B + \frac{C_2}{R_B}=C_3 R_B + \frac{C_4}{R_B}\\
 (-C_3+\frac{C_4}{R_c^2})\sigma_B=-C_5 \sigma_c \\
 (E_0+\frac{C_2}{R_B^2})\sigma_m=(-C_3+\frac{C_4}{R_B^2})\sigma_B
 \end{split}
 \end{equation}

\noindent which can be solved for $C_2$, $C_3$, $C_4$ and $C_5$ as functions of $\sigma_c$, $\sigma_B$, $\sigma_m$, $R_c$ and $R_B$.
If we impose that $C_2(\sigma_c, \sigma_B, \sigma_m, R_c, R_B)=0$, it is not difficult to see from Eq. \ref{solution} that the electric field outside the area where the magnetic field is applied will be equal in magnitude and direction to the uniform electric field at infinity. This constitutes the base of the cloaking condition: the electric field distribution where the magnetic field is applied will be concealed and this region will be invisible in terms of electric currents. If we let $C_2(\sigma_c, \sigma_B, \sigma_m, R_c, R_B)=0$ in the solution of Eq. \ref{sistema de 4x4} for $C_2$ we obtain the cloaking condition:

  \begin{equation} \label{cloaking_condition_SOM}
 \sigma_m= \frac{R_B^2(\sigma_c+\sigma_B)+R_c^2(\sigma_c-\sigma_B)}{R_B^2(\sigma_c+\sigma_B)+R_c^2(\sigma_B-\sigma_c)} \sigma_B
  \end{equation}

This condition for the externally driven current cloak can be reduced to a particular case already published \cite{Ma2014}.

If we considered $\sigma_c=0$, Eq. \ref{cloaking_condition_SOM} reduces to:

\begin{equation} \label{cloaking_condicion_Ma}
 \sigma_m= \frac{R_B^2-R_c^2}{R_B^2+R_c^2} \sigma_B
  \end{equation}

This condition was obtained by Yungui Ma {\it et al.} working with a bifunctional cloak \cite{Ma2014} and corresponds to a core consisting in a void ($\sigma_c = 0$). Here $\sigma_B>\sigma_m$ and all parameters have to be fixed to properly fulfill the cloaking condition. Our cloak allows us to apply an external field to “tune” the cloaking state instead of permanently modifying the material.

It is worth noting that in our case $\sigma_c$ must be larger than $\sigma_B$ in order to obtain the cloak. If $\sigma_c$ is smaller than $\sigma_B$, the $\sigma_m / \sigma_B$ ratio in Eq. \ref{cloaking_condition_SOM} (looking at the right-hand side of the equation after dividing both sides by $\sigma_B$) would be smaller than $1$ , letting the conductivity of the region where the magnetic field is applied (excepting the inclusion) be larger than that of the external material. This scenario is not possible using a magneto-resistive material and a magnetic field as described above, but it is part of the solution and a perfect current cloak can be built with such specs.

\subsection{Proximity of the boundaries}

\begin{figure}[b]
\begin{center}
\includegraphics[width=0.7\linewidth]{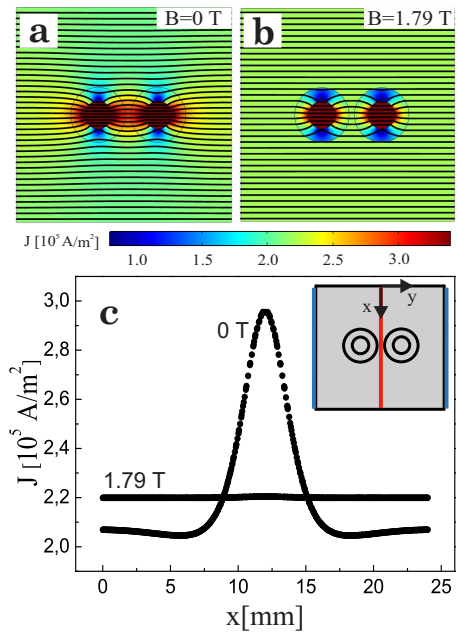}
\caption{\label{FIG:cloakstogether} Cohabiting cloaks. (a,b) Current density lines for a 1-mm thickness polycrystalline Bi matrix with two stainless steel inclusions of $R_c =$ 1.6 mm radii (inner circles), and magnetic fields applied within concentric regions of radii $R_B =$ 3.2 mm (outer circles). (a) and (b) correspond to magnetic fields of $0$ T and $1.79$ T, respectively. (c) Current density values plotted along a vertical line tangent to the area of application of the magnetic field (see inset).}
\end{center}
\end{figure}

Fig. \ref{FIG:boundariesproximity} (a,b) shows a Finite Elements simulation of the current density lines along a 1-mm thickness polycrystalline Bi matrix with a stainless steel inclusion of $R_c =$ 1.6 mm radius (inner circle), and a magnetic field applied within a region of radius $R_B =$ 3.2 mm (outer circle). The positions of the upper and lower boundaries were changed allowing a separation of $1/7$ its radius from the device. The presence of the highly conductive inclusion in the absence of magnetic field deforms current density lines (Fig. \ref{FIG:boundariesproximity} (a)); on the other hand, when we applied $B_{cloak}$ = $1.79$ T in a region of radius 3.2 mm concentric with the inclusion (Fig. \ref{FIG:boundariesproximity} (b)) the inhomogeneities are concealed and we obtain a perfect current cloak. Fig. \ref{FIG:boundariesproximity} (c) allows us to see in detail what happens to the current density along a vertical line running from the top to the bottom of the sample (red line presented in the inset).

The cloak is obtained at the same value of $B_{cloak}$, making confined scenarios suitable to apply Eq. \ref{cloaking_condition_SOM} without any correction. This let us conclude that, as long as the electric field is uniform in the absence of magnetic field or inclusion, the proximity of the boundaries is not going to affect the cloaking state.

\subsection{Cloaks coexisting together}

Fig. \ref{FIG:cloakstogether} (a,b) shows the current density lines along a 1-mm thickness polycrystalline Bi matrix with two stainless steel inclusions of $R_c =$ 1.6 mm radii (inner circles), and magnetic fields applied within concentric regions of radii $R_B =$ 3.2 mm (outer circles). The current density field is deformed by the highly conductive inclusions at zero magnetic field in the expected way (Fig. \ref{FIG:cloakstogether} (a)); however, when a magnetic field of induction $B_{cloak}$ = $1.79$ T is applied in a region of radius 3.2 mm around both inclusions (Fig. \ref{FIG:cloakstogether} (b)) we obtain a perfect concealing of inhomogeneities. Fig. \ref{FIG:cloakstogether} (c) exhibits the changes of the current density along a vertical line running from the top to the bottom of the sample (red line presented in the inset), emphasizing the attainment of the cloaks at $B$ = $1.79$ T.

If the magnetic field in one of the cloaks (or the area where it is applied) is not correct, both cloaks are going to be out of the cloaking state. The conditions stated in Eq. \ref{cloaking_condition_SOM} have to be individually satisfied so as to achieve a collective result. We examined several configurations changing position, number, and radii of the cores; and, in all cases, we obtained a concealing of inhomogeneities when all cloaks were perfectly tuned.

\raggedbottom
\pagebreak

\end{document}